\documentclass[twocolumn,showpacs,aps,prb]{revtex4}
\usepackage{graphicx}
\usepackage{dcolumn}
\usepackage{bm}
\usepackage{color}
\usepackage{marvosym}

\begin{document}
\preprint{Physical Review B} %
\title{Spin-dependent Rotating Wigner Molecules in Quantum dots}
\author{Zhensheng Dai}
\affiliation{%
Department of Physics, and Center for Quantum Information, Tsinghua
University, Beijing 100084, People's Republic of China}
\author{Jia-Lin Zhu}
\email[Electronic address: ]{zjl-dmp@tsinghua.edu.cn}
\affiliation{%
Department of Physics, and Center for Quantum Information, Tsinghua
University, Beijing 100084, People's Republic of China}
\author{Ning Yang}
\affiliation{%
Department of Physics, and Center for Quantum Information, Tsinghua
University, Beijing 100084, People's Republic of China}
\author{Yuquan Wang}
\affiliation{%
Department of Physics, and Center for Quantum Information, Tsinghua
University, Beijing 100084, People's Republic of China}

\date{\today}

\begin{abstract}
The spin-dependent trial wave functions with rotational symmetry are introduced
to describe rotating Wigner molecular states with spin degree of freedom in
four- and five-electron quantum dots under magnetic fields. The functions are
constructed with unrestricted Hartree-Fock orbits and projection technique in
long-range interaction limit. They highly overlap with the exact-diagonalized
ones and give the accurate energies in strong fields. The zero points, i.e.
vortices of the functions have straightforward relations to the angular momenta
of the states. The functions with different total spins automatically satisfy
the angular momentum transition rules with the increase of magnetic fields
and explicitly show magnetic couplings and characteristic oscillations with
respect to the angular momenta. Based on the functions, it is
demonstrated that the entanglement entropies of electrons depend on the
z-component of total spin and rise with the increase of angular momenta.
\end{abstract}
\maketitle%
\section{Introduction}
The quantum dot systems, especially the few-electron dots with
precise control of the particle number, have been extensively
studied in recent years due to their abundant application potential
and theoretical meaning. Experimentally, these systems are
investigated as candidates for future quantum electronics,
spintronics\cite{Wolf2001} and quantum information
devices\cite{Loss1998, Cerletti2005}. Theoretically, such
two-dimensional (2D) confined systems with or without external
fields have significant differences with the extent ones. In strong
magnetic fields, the 2D electron gas with short-range interaction
which exhibits the fractional quantum Hall effect has been
understood based on the Laughlin and composite fermion wave
functions.\cite{Laughlin1983, Jain1989} It is also anticipated that
there will be the transition from the quantum liquid to the Wigner
crystal in the magnetic fields corresponding to very small filling
factors.\cite{Yang2001, Ye2002, Mandal2003, He2005} In quantum dots,
it has been demonstrated that similar transition is much easier than
that in extent systems due to the intrinsic long-range
interactions.\cite{Yannouleas2003, Yannouleas2004} And both the
exact-diagonalization method (ED) and the analytic theory confirm
that the crystal states in quantum dots are not static but rotating
ones with certain angular momenta, which are often referred to
rotating Wigner molecules (RWMs).

In most investigations on few-electron states in quantum dots in
strong magnetic fields, the spins of electrons are considered
full-polarized and it simplifies the analytical theory and
computational demands. However, in recent experiments, it has been
possible to fabricate and study the quantum dots with negligible
Zeeman splitting.\cite{Ellenberger2006, Salis2001} Then the full
considerations of spin degree of freedom are needed and pose
challenge to theorists. Although it has been demonstrated that the
spin degree of freedom does not change the cystallization process in
quantum dots, it will still bring new physical phenomena to the
system.

In this work, we will introduce the four- and five-electron trial wave
functions for the spin-dependent RWMs. Noticing that the similar four-electron
wave function has been proposed recently by Jain et al\cite{Jeon2007} and their
results are consonant with our four-electron case, we will demonstrate that
the functions highly agree with the ED results and concentrate on the different
quantum characters between four and five electron quantum dots without the
Zeeman splitting in strong magnetic fields. The remainder of the paper is
organized as follows. The descriptions of the trial wave functions are
introduced in Sec.II, the discussion of the accuracy of the functions and their
quantum characters, such as the vortex numbers, magnetic couplings and
entanglements between electrons are presented in Sec.III followed by a summary
in Sec.IV.

\section{Wave Functions}
In strong magnetic fields, unrestricted Hartree-Fock(UHF) orbits
describing localized electrons can be approximated by displaced
Gaussian functions\cite{Yannouleas2002}
\begin{equation}
u(z,Z_j)\!\!=\!\!\frac{1}{\sqrt{\pi}\lambda}\exp\!\!\left(-\frac{|z-Z_j|^2}{2\lambda^2}\right)
\!\!\exp\!\!\left(\frac{zZ_j^*-z^*Z_j}{4l_B^2}\right)
\end{equation}
where $\lambda=\sqrt{\hbar/m^*\Omega}$ ,
$\Omega=\sqrt{\omega_0^2+\omega_c^2/4}$ , $l_B=\sqrt{\hbar c/eB}$.
$z=x-\mathrm{i}y$ is the complex coordinate of an electron.
$Z_j=X_j+\mathrm{i}Y_j$ are the centers of the Gaussians and set to
the positions of the equilibrium positions of electrons. The phase
factor ensures the gauge invariance of each localized orbit. When
$\omega_0=0$ or $\omega_c\gg\omega_0$, i.e. $\lambda=\sqrt{2}l_B$,
$u(z,Z_j)$ can be expanded to the first landau level
\begin{equation}\label{EQ:Expand}
u(z,Z_j)=\sum_{l=0}^{\infty}{c_l(Z_j)\varphi_l(z)}
\end{equation}
where
$c_l(Z_j)=(Z_j^*/\lambda)^l\exp(-|Z_j|^2/2\lambda^2)/\sqrt{l!}$.

For a dot with $N_1$ spin-up and $N_2$ spin-down electrons
($N_1+N_2=N$), if the electrons form a RWM in a single ring
configuration, we can construct $C_N^{N_1}$ many-body bases
$|Z_{j_1}^\uparrow,...,
Z_{j_{N_1}}^\uparrow,Z_{j_{N_1+1}}^\downarrow,...,Z_{j_N}^\downarrow\rangle$
with z-component $S_z=(N_1-N_2)/2$ from the localized UHF orbits
centered at positions $Z_1\sim Z_N$ in clockwise, i.e.
$Z_j=\exp(\mathrm{i}2\pi(1-j)/N)$. $\uparrow$ ($\downarrow$)
represents the spin of the electron and each position can be only
occupied by one particle with a certain spin.

Taking the four-electron case with $S_z=0$ as an example, the
many-particle bases are
\begin{equation} \left(
\begin{array}{c}
|1\rangle\\|2\rangle\\|3\rangle\\|4\rangle\\|5\rangle\\|6\rangle
\end{array}
\right)=\left(
\begin{array}{c}
|Z_1^\uparrow,Z_2^\uparrow,Z_3^\downarrow,Z_4^\downarrow\rangle\\
|Z_1^\uparrow,Z_3^\uparrow,Z_2^\downarrow,Z_4^\downarrow\rangle\\
|Z_1^\uparrow,Z_4^\uparrow,Z_2^\downarrow,Z_3^\downarrow\rangle\\
|Z_2^\uparrow,Z_3^\uparrow,Z_1^\downarrow,Z_4^\downarrow\rangle\\
|Z_2^\uparrow,Z_4^\uparrow,Z_1^\downarrow,Z_3^\downarrow\rangle\\
|Z_3^\uparrow,Z_4^\uparrow,Z_1^\downarrow,Z_2^\downarrow\rangle
\end{array}
\right).
\end{equation}

In the second quantization scheme, the N-electron Hamiltonian can be
written as
\begin{equation}
H=\sum_i{\epsilon_i
a_i^+a_i}+\frac{1}{2}\sum_{ijkl}{V_{ijkl}a_i^+a_j^+a_la_k}.
\end{equation}
The orbital energies and the direct Coulomb integrals of UHF orbits
which are same for all bases and can be eliminated from the diagonal
elements of $H$. If we assume that the exchange integrals between
the neighbor electrons and next neighbors are $V_{ijji}=v_1$ and
$V_{ikki}=v_2$, the Hamiltonian will have a simple form. For four
electrons, it is
\begin{equation}\label{EQ:Hami4E}
H=\left(
    \begin{array}{cccccc}
      -2v_1 & v_1   & -v_2  & -v_2  & v_1   & 0 \\
      v_1   & -2v_2 & v_1   & v_1   & 0     & v_1 \\
      -v_2  & v_1   & -2v_1 & 0     & v_1   & -v_2 \\
      -v_2  & v_1   & 0     & -2v_1 & v_1   & -v_2 \\
      v_1   & 0     & v_1   & v_1   & -2v_2 & v_1 \\
      0     & v_1   & -v_2  & -v_2  & v_1   & -2v_1 \\
    \end{array}
  \right).
\end{equation}
The eigenstates of $(H,S^2,S_z)$ can be obtained by the
diagonalization of $H$. The eigenvalues and eigenstates of
Eq.(\ref{EQ:Hami4E}) are
\begin{equation}\label{EQ:EigenEV}
\begin{array}{lll}
S=0 & \left\{\begin{array}{l}
E_1=2v_1-2v_2\\E_2=2v_2-2v_1\end{array}\right.&
\begin{array}{l}\psi_1:(1, 2, 1, 1, 2, 1)\\\psi_2:(1, 0, -1, -1, 0,
1)\end{array}\\
S=1 & \left\{\begin{array}{l}
E_3=-2v_1\\E_4=-2v_1\\E_5=-2v_2\end{array}\right. &
\begin{array}{l}\psi_3:(0, 0, -1, 1, 0, 0)\\\psi_4:(-1, 0, 0, 0, 0,
1)\\\psi_5:(0, -1, 0, 0, 1, 0)\end{array}\\
S=2 & \hspace{2.8ex}E_6=-4v_1-2v_2 &\hspace{0.6ex} \psi_6:(1, -1, 1,
1, -1, 1)
\end{array}
\end{equation}
\begin{widetext}
Using Eq.(\ref{EQ:Expand}), the many-particle bases can be expanded
to the first Landau level as

\begin{eqnarray}
|Z_{j_1}^\uparrow,...,Z_{j_{N_1}}^\uparrow,Z_{j_{N_1+1}}^\downarrow,...,Z_{j_N}^\downarrow\rangle
=\!\!\!\sum_{l_1,l_2,...,l_N=0}^{\infty}\!\!\!{c_{l_1}(Z_{j_1})c_{l_2}(Z_{j_2})...c_{l_N}(Z_{j_N})}
{|l_1^\uparrow,...,l_{N_1}^\uparrow,l_{N_1+1}^\downarrow,...,l_{N}^\downarrow\rangle}.
\end{eqnarray}
These many-particle bases are breaking symmetrical and contain the
components of all angular momenta. The component of angular momentum
$L$ can be obtained by projection operator
technique\cite{Yannouleas2002}
\begin{eqnarray}
|Z_{j_1}^\uparrow,...,Z_{j_{N_1}}^\uparrow,Z_{j_{N_1+1}}^\downarrow,...,Z_{j_N}^\downarrow\rangle_L
=\!\!\!\sum_{0\leq l_1<l_2<...<l_{N_1} \atop 0\leq
l_{N_1+1}<...<l_{N}}^{l_1+l_2+...+l_N=|L|}
\!\!\!{\det[c_{l_1}(Z_{j_1}),c_{l_2}(Z_{j_2}),...,c_{l_{N_1}}(Z_{j_{N_1}})]}\nonumber\\
{\det[c_{l_{N_1+1}}(Z_{j_{N_1+1}}),c_{l_{N_1+2}}(Z_{j_{N_1+2}}),...,c_{l_{N}}(Z_{j_{N}})]}
{|l_1^\uparrow,...,l_{N_1}^\uparrow,l_{N_1+1}^\downarrow,...,l_{N}^\downarrow\rangle}.
\end{eqnarray}
\end{widetext}
For four electrons again, they are (up to a constant)
\begin{eqnarray}\label{EQ:Cof4e}
|k\rangle_L\!\!&=&\nonumber\\
&&\!\!\sum_{0\leq l_1<l_2,0\leq
l_3<l_4}^{l_1+l_2+l_3+l_4=|L|}{\!\!\left(\prod_{j=1}^4{l_j!}\right)^{-1/2}
\!\!\!\!\!\!\!\!\left(\mathrm{i}^{kl_2}-\mathrm{i}^{kl_1}\right)
\left(\mathrm{i}^{kl_4}-\mathrm{i}^{kl_3}\right)}\nonumber\\
&&\!\!\times(-1)^{(l_3+l_4)/k}|l_1^\uparrow,l_2^\uparrow,l_3^\downarrow,l_4^\downarrow\rangle
\end{eqnarray}
where $k=1,2$, and
\begin{equation}\label{EQ:Factor}
\begin{array}{ccr}
|3\rangle_L&=&-\exp(\mathrm{i}6\pi|L|/4)|1\rangle_L\\
|4\rangle_L&=&-\exp(\mathrm{i}2\pi|L|/4)|1\rangle_L\\
|5\rangle_L&=&-\exp(\mathrm{i}2\pi|L|/4)|2\rangle_L\\
|6\rangle_L&=&\exp(\mathrm{i}4\pi|L|/4)|1\rangle_L
\end{array}
\end{equation}
due to the $\pi/2$ rotational symmetry of the positions $Z_1\sim Z_4$. Then six
eigenstates with conserved angular momentum can be written as the linear
combinations of $|1\rangle_L$ and $|2\rangle_L$ using the relations in
Eq.(\ref{EQ:Factor}). Since the phase differences in Eq.(\ref{EQ:Factor}), for
the states with a certain total spin, only those with proper $L$ have
nonvanishing combination coefficients. The allowable $L$ and corresponding
coefficients for the eigenstates are listed in Tab.\ref{TAB:MC4E}. It can be
also recognized that $|1\rangle_L$ and $|2\rangle_L$ imply respectively the
ferromagnetic and anti-ferromagnetic coupling between electrons. This fact will
be important in following discussion of the properties of spin-dependent RWMs.

For five-electron case with $S_z=0.5$, repeat the same scheme and
again for the two bases
$|Z_1^\uparrow,Z_2^\uparrow,Z_3^\uparrow,Z_4^\downarrow,Z_5^\downarrow\rangle$
and
$|Z_1^\uparrow,Z_2^\uparrow,Z_4^\uparrow,Z_3^\downarrow,Z_5^\downarrow\rangle$
which are ferromagnetic and ferrimagnetic, the components with
angular momentum $L$ are respectively
\begin{eqnarray}\label{EQ:5eBasis1}
&&|Z_1^\uparrow,Z_2^\uparrow,Z_3^\uparrow,Z_4^\downarrow,Z_5^\downarrow\rangle_L\!\!=\nonumber\\
&&\!\!\sum_{0\leq l_1<l_2<l_3,0\leq
l_4<l_5}^{l_1+l_2+l_3+l_4+l_5=|L|}
{\!\!\left(\prod_{j=1}^5{l_j!}\right)^{-1/2}}{\!\!\!\!\!\!\!\!\left(\delta^{l_2}-\delta^{l_1}\right)
\left(\delta^{l_3}-\delta^{l_2}\right)}\nonumber\\
&&\!\!\times\left(\delta^{l_3}-\delta^{l_1}\right)\left(\delta^{l_5}-\delta^{l_4}\right)\delta^{3(l_4+l_5)}|l_1^\uparrow,l_2^\uparrow,l_3^\uparrow,l_4^\downarrow,l_5^\downarrow\rangle
\end{eqnarray}
\begin{eqnarray}\label{EQ:5eBasis2}
&&|Z_1^\uparrow,Z_2^\uparrow,Z_4^\uparrow,Z_3^\downarrow,Z_5^\downarrow\rangle_L\!\!=\nonumber\\
&&\!\!\sum_{0\leq l_1<l_2<l_3,0\leq
l_4<l_5}^{l_1+l_2+l_3+l_4+l_5=|L|}
{\!\!\left(\prod_{j=1}^5{l_j!}\right)^{-1/2}}{\!\!\!\!\!\!\!\!\left(\delta^{l_2}-\delta^{l_1}\right)
\left(\delta^{l_3}-\delta^{l_2}\right)}\nonumber\\
&&\!\!\times\left(\delta^{l_3}-\delta^{l_1}\right)\left(\delta^{2l_5}-\delta^{2l_4}\right)\left(\delta^{l_1}+\delta^{l_2}+\delta^{l_3}\right)\nonumber\\
&&\!\!\times\delta^{2(l_4+l_5)}|l_1^\uparrow,l_2^\uparrow,l_3^\uparrow,l_4^\downarrow,l_5^\downarrow\rangle
\end{eqnarray}
where $\delta=\exp(\mathrm{i}\frac{2}{5}\pi)$. And at this time,
there are ten eigenstates as the linear combinations of the two
bases. See Tab.\ref{TAB:MC5E} for corresponding coefficients.

\begin{table}[!h]
\caption{Eigenvalues, angular momenta, total spins, combination
coefficients and magnetic couplings (MC) of the trial wave functions
for four-electron spin-dependent RWMs}\label{TAB:MC4E}
\begin{tabular}{|c|c|c|c|c|c|c|}
\hline &$E$&$|L|$&$S$&$C_1$&$C_2$&MC\\
\hline $\psi_1$&$2v_1-2v_2$&$4n+2$&0&1&1&1\\
\hline $\psi_2$&$2v_2-2v_1$&$4n$&0&1&0&-1\\
\hline $\psi_3$&$-2v_1$&$4n\pm1$&1&1&0&-1\\
\hline $\psi_4$&$-2v_1$&$4n\pm1$&1&$\mp\mathrm{i}$&0&-1\\
\hline $\psi_5$&$-2v_2$&$4n$&1&0&1&1\\
\hline $\psi_6$&$-4v_1-2v_2$&$4n+2$&2&1&-0.5&0\\ \hline
\end{tabular}
\end{table}
\begin{table}[!h]
\caption{Eigenvalues, angular momenta, total spins, combination
coefficients and magnetic couplings (MC) of the trial wave functions
for five-electron spin-dependent RWMs}\label{TAB:MC5E}
\begin{tabular}{|c|c|c|c|c|c|c|}
\hline &$E$&$|L|$&$S$&$C_1$&$C_2$&MC\\
\hline $\psi_1$&$-(v_1+v_2)$&$5n$&0.5&1&1&0\\
\hline $\psi_{2,3}$&$-(v_1+v_2)-\sqrt{5}(v_1-v_2)$&$5n\pm2$&0.5&1&$\delta_1$&-1\\
\hline $\psi_{4,5}$&$-(v_1+v_2)+\sqrt{5}(v_1-v_2)$&$5n\pm1$&0.5&$\delta_2$&1&1\\
\hline $\psi_{6,7}$&$-\frac{5}{2}(v_1+v_2)-\frac{\sqrt{5}}{2}(v_1-v_2)$&$5n\pm1$&1.5&1&$\delta_3$&-1\\
\hline $\psi_{8,9}$&$-\frac{5}{2}(v_1+v_2)+\frac{\sqrt{5}}{2}(v_1-v_2)$&$5n\pm2$&1.5&$\delta_4$&1&1\\
\hline $\psi_{10}$&$-5(v_1+v_2)$&$5n$&2.5&1&-1&0\\ \hline
\end{tabular}
\indent\flushleft
$\delta_1=\Delta\exp(\mp\mathrm{i}\frac{3}{5}\pi)$,
$\delta_2=\Delta\exp(\mp\mathrm{i}\frac{1}{5}\pi)$,
$\delta_3=\Delta\exp(\mp\mathrm{i}\frac{4}{5}\pi)$,
$\delta_4=\Delta\exp(\mp\mathrm{i}\frac{2}{5}\pi)$,
$\Delta=0.381966$.
\end{table}

It is worthwhile to point out that the projection lead to the fact
that not all the eigenstates with restored rotational symmetry are
still orthogonal even if the breaking symmetrical ones do. It can be
seen in Tab.\ref{TAB:MC4E} that only five of six eigenstates are
orthogonal. $\psi_3$ and $\psi_4$ are the same state up to a phase
factor. The situation is similar in five-electron case as shown in
Tab.\ref{TAB:MC5E}.

The scheme for constructing the trial functions with $S_z\neq0
(0.5)$ is same. Because in the following of the paper we will mainly
discuss the RWMs with $S_z=0 (0.5)$, the other wave functions are
not shown here and will be attached in appendix.

Having got the eigenstates, we can evaluate their actual energies by
considering the Hamiltonian of quantum dots with parabolic
confinement and exact interactions as
\begin{equation}
H\!\!=\!\!\sum_{i=1}^N{\!\left(\frac{(\hat{P}_i\!+\!e\vec{A})^2}{2m_e^*}+\frac{1}{2}m_e^*\omega_0^2r_i^2\right)}\!\!+\!\!\sum_{i<j}{\frac{e^2}{4\pi\epsilon|\vec{r}_i-\vec{r}_j|}}
\end{equation}
The energy expectation value
${\langle\psi|H|\psi\rangle}/{\langle\psi|\psi\rangle}$ of the trial
function with angular momentum $L$ are
\begin{equation}\label{EQ:ExV}
E_{dot}(B)=\frac{N}{2}\hbar\omega_c+(L+N)\hbar\frac{\omega_0^2}{\omega_c}+\frac{l_0}{l_B}\langle
V\rangle
\end{equation}
where $\langle V\rangle$ is the expectation value of interaction
energy between electrons evaluated with $l_B$ equals to a certain
length $l_0$.

Noticing that the overlap of a single-particle orbit
$\varphi_l(z)$ in a quantum dot with the first Landau level
$\varphi_l(z)_{\omega_0=0}$ is
$[2\sqrt{2}\lambda\l_B/(\lambda^2+2l_B^2)]^{|l|+1}$, we can also
evaluate the overlap between the trial function with angular
momentum $L$ and the exact-diagonalized one as
\begin{equation}
O_L=\left(\frac{2\sqrt{2}\lambda\l_B}{\lambda^2+2l_B^2}\right)^{|L|+N}\hspace{-5ex}\cdot(C_{T},C_{E})
\end{equation}
where $(C_T,C_E)$ represents the inner product between the
determinant coefficient vectors of trial function and that of
exact-diagonalized one.

In order to study the quantum correlations in the trial functions,
we can calculate the von Neumann entropies\cite{Paskauskas2001}
between an electron and the other part of the system as
\begin{equation}
S=-\mathrm{tr}(\rho^f\log_2\rho^f)
\end{equation}
where $\rho_{\mu,\nu}^f=\langle\psi|a_\mu^+a_\nu|\psi\rangle$ is
the single-particle reduced density matrix.

\section{Disscussion}
\subsection{Energies, overlaps and vortices of the functions}
\begin{figure}[ht]
\includegraphics*[width=0.42\textwidth]{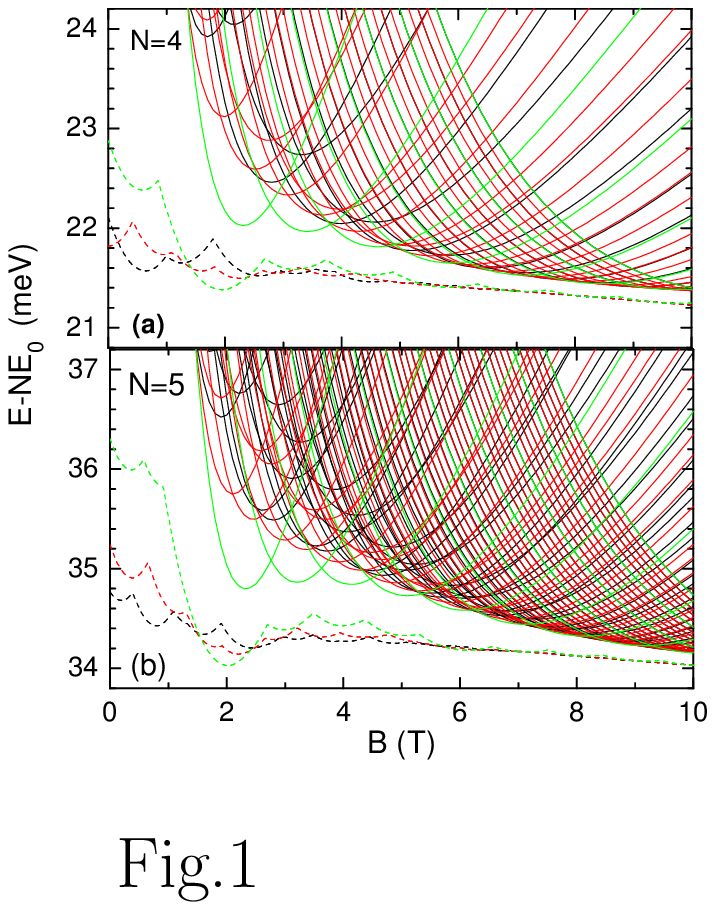}
\caption{\label{FIG:Energy}(Color online) Energy expectation values
of the trial wave functions (solid lines) and the energy of the
lowest states calculated by ED (dashed lines) for four (a) and five
electrons (b) with total spins from the minimum to maximum values
indicated by black, red and green lines, respectively. N times of
the energy of the first Landau level have been subtracted from the
total energy.}
\end{figure}
The energy expectation values of the trial functions obtained from
Eq.(\ref{EQ:ExV}) as function of magnetic field are plotted in
Fig.\ref{FIG:Energy}. In the same plot, we also present the energies
of lowest states with different total spins calculated by ED. The ED
calculations are executed beyond the lowest Landau level
approximation and the confinement strength is set to 2meV. Then we
employ the results of ED as a criterion to examine the accuracy of
the trial functions. It can be seen in the plot that the results for
four electrons obtained from the trial wave functions agree well
with the ED ones in the magnetic field stronger than 5.7T with only
an increase of the total energies lower than 0.25meV ($0.6\%$ of the
total energy at $B=5.7T$). For five electrons, the difference is
lower than 0.398meV (still $0.6\%$ of the total energy) when the
field is stronger than 6T.

With the change of the field, both trial functions and ED exhibit
accordant angular momentum transitions for the states with different
total spins. In fact the allowable angular momenta of the trial
functions listed in Tab.\ref{TAB:MC4E} and \ref{TAB:MC5E} are just
the angular momenta existing in the transitions identified by ED.
When the magnetic field is strong enough, the trial functions can
give correct energy sequence of different spin states and the values
of the field where the transitions take place. In smaller fields,
the full-polarized rotating Wigner molecules always have lower
energies than the ones with other total spins if the states in
quantum dots are described by the trial functions. However, the ED
demonstrates that the states in small fields are actually liquidlike
and the states which are not full-polarized can have lower energies.
Then the trial functions cannot describe them correctly. It should
be noticed that because the field ranges for liquid-to-crystal
transition depend on the confinements of quantum dots, the trial
functions can be expected to give better agreement with ED ones in
smaller fields for the dot with weaker confinement.

\begin{figure}[h]
\includegraphics*[width=0.4\textwidth]{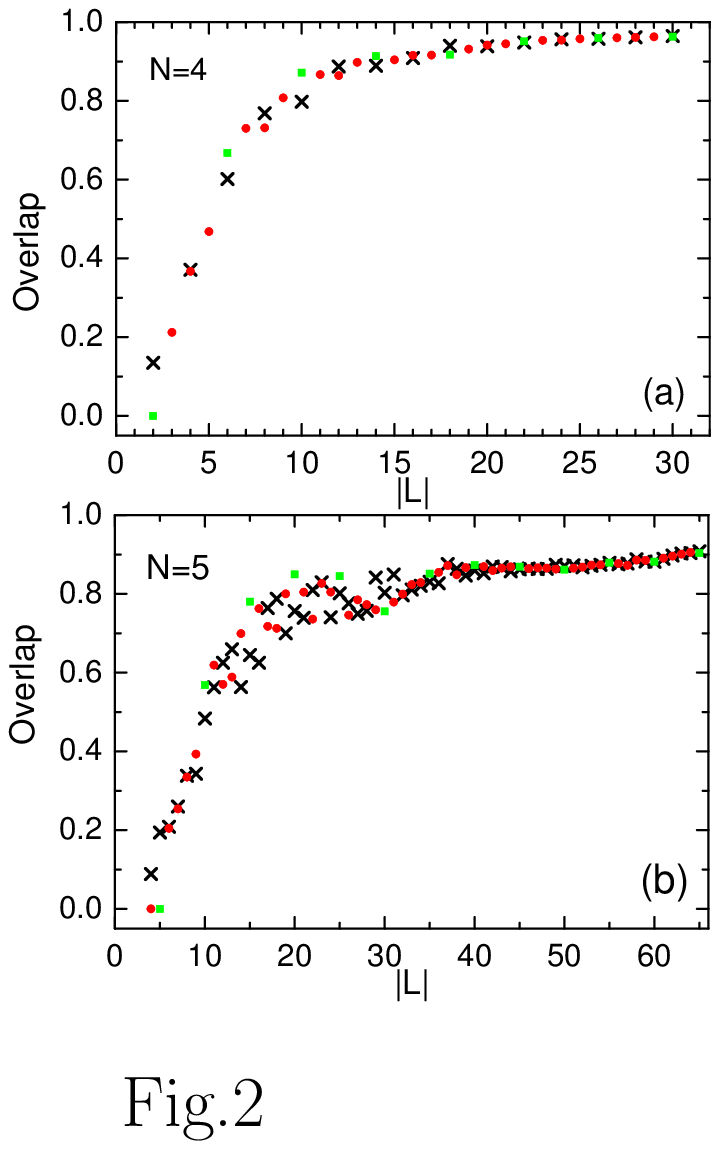}
\caption{\label{FIG:Overlap}(Color online) Overlaps between the
trial wave functions and ED ones for four (a) and five electrons (b)
as function of angular momentum. $\times$, $\bullet$ and
{\Rectsteel} correspond to the states with non-, partial- to
full-polarized spins, respectively.}
\end{figure}

\begin{figure*}[ht]
\includegraphics*[width=0.8\textwidth]{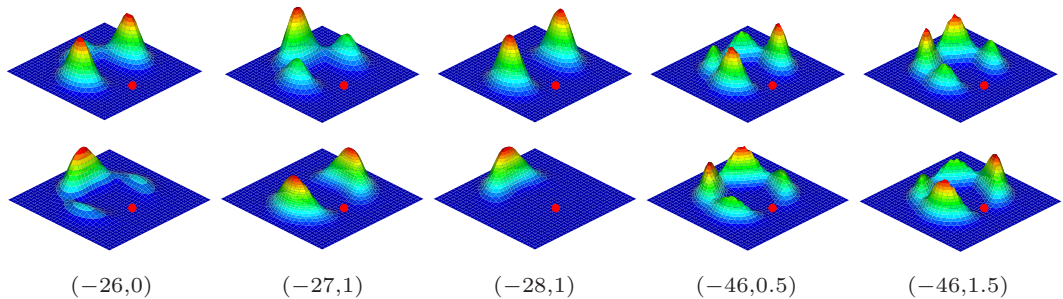}
\caption{\label{FIG:Spin}(Color online) Spin-dependent conditional
probability densities of four-electron states with $(L,S)=(-26,0)$
$(-27,1)$ and (-28,1) and five-electron states $(-46,0.5)$ and
$(-46,1.5)$ in QDs.The panels in the first(second) row show the
distributions of the spin-down(up) electron. The red dots indicate
the positions of the fixed spin-up electron.}
\end{figure*}

Besides the energy expectation values, we also evaluate the overlaps between
the trial functions and the ED ones. As shown in Fig.\ref{FIG:Overlap}, the higher
overlaps convinced us that the trial functions can describe the states with
larger angular momenta in quantum dots, i.e. the lowest states in strong field.
The zero points, i.e. vortices of the many-body wave functions can reflect the
charge distributions,\cite{Saarikoski2004} so we also inspect the vortex
structures of the trial functions. The vortex number of the function just
equals to the highest angular momentum occupied by the single particle. For
four electron non-full-polarized states with $S_z=0$, the vortex number
generally equals to $|L|-1$ as the function contains the component corresponds
to the angular momentum distribution $\{l_1,l_2,l_3,l_4\}=\{0,1,0,|L|-1\}$ or
$\{l_1,l_2,l_3,l_4\}=\{0,|L|-1,0,1\}$. An exception is the case when
$|L|-1=4n$, where $n$ is an arbitrary integer. From Eq.(\ref{EQ:Cof4e}) it can
be found that the coefficient of such component is zero, so the vortex numbers
of these states are $|L|-2$. For the full-polarized states, the vortex numbers
are $|L|-3$ due to the Pauli exclusion principle. We also compare the vortex
structures of the trial functions with that of ED ones. Although the angular
momentum components of the trial functions are restricted in the first Landau
level, the vortex number in the scope which contains the majority of the
electron charge can agree well with the ED ones. Of course, due to the fact
that our trial functions mainly reflect the long-range limit characters of the
electrons in quantum dots, the vortex distributions are more dispersed and
different from the ED ones.

For the energies, overlaps and vortices, the magnetic field or
angular momentum beyond which the trial functions can get agreement
with the ED ones for five electrons is larger than that for
four-electron case. This is because that the transition from liquid
to crystal states needs larger fields for the dots with larger
particle numbers.

\subsection{Spin correlations of RWMs}
It has been pointed out in Sec.II that the many-body bases
$|1\rangle_L$ and $|2\rangle_L$ for four-electron imply two species
of magnetic coupling existing among electrons, respectively. Then
the spin-dependent RWMs may also have corresponding properties since
they are linear combinations of the two bases.

As listed in Tab.\ref{TAB:MC4E}, we have indicated the
ferromagnetic, anti-ferromagnetic couplings and no specific coupling
with number -1, 1 and 0 respectively. There are four-electron states
like $\psi_{2\sim5}$ which only contain the component $|1\rangle_L$
or $|2\rangle_L$ and of course they should have the corresponding type of
magnetic coupling. There are also states like $\psi_1$ and $\psi_5$
containing both of the components. The full-polarized states
$\psi_6$ should not have any specific magnetic coupling because the
spin parts of the wave functions can be separated from the spatial
parts. Then the states $\psi_1$ should have residual
anti-ferromagnetic coupling because the ratio of the two components
in them is different from that in $\psi_6$.

We also calculate the conditional probability densities (CPDs) of
finding electrons with a certain species of spin to explicitly
identify the two kind of magnetic couplings. As shown in
Fig.\ref{FIG:Spin}, if a spin-up electron is fixed, the remainder
spin-up and -down electrons of four-electron state with $L=-27$,
$S=1$ will be clearly at the neighbor and opposite positions,
respectively. There are indeed the probability to find spin-down
electrons at the neighbor positions. This is because the spin-down
electrons have equal probability to stay at two remainder position
if the spin-up electron is at left or right neighbor of the fixed
ones. Such CPDs just reflect the ferromagnetic coupling between
electrons. Then the other CPDs as the case of the state with
$L=-28$, $S=1$ are the reflection of anti-ferromagnetic coupling. In
these two examples, the states only contain the single component
$|1\rangle_L$ or $|2\rangle_L$ respectively. The case of the state
with $L=-26$, $S=0$ is also anti-ferromagnetic as the states with
$L=-28$, $S=1$. However their CPDs have a slight difference since
the state with $L=-26$, $S=0$ is just a sample of $\psi_1$ which
contain both components $|1\rangle_L$ and $|2\rangle_L$.

Similar to the four-electron case, the five-electron bases
Eq.(\ref{EQ:5eBasis1}) and Eq.(\ref{EQ:5eBasis2}) are ferromagnetic and
ferrimagnetic respectively. Then the five-electron spin-dependent RWMs also
have corresponding magnetic coupling as listed in Tab.\ref{TAB:MC5E}, where
$-1$,1 and 0 respectively represent ferromagnetic, ferrimagnetic and no
specific couplings. Different from the four-electron case, all of the
five-electron RWMs contain both components of the two bases, they can only have
the residual magnetic coupling if the ratios of the two components are
different from that in full-polarized states $\psi_{10}$. The states like
$\psi_1$ which have same ratio as $\psi_{10}$ also have no specific magnetic
coupling. In Fig.\ref{FIG:Spin} we also show the CPDs of two five-electron
states with $L=-46,S=0.5$ and $S=1.5$ which respectively reflect ferrimagnetic
and ferromagnetic couplings.

Before the finish of the subsection, it is worthwhile to point out that there
are regular magnetic coupling oscillations for spin-dependent RWMs with respect
to the angular momentum which are in accordance with the electron molecules
theory.\cite{Maksym2000} It can be identified in Tab.\ref{TAB:MC4E} and
\ref{TAB:MC5E} that with the increase of the angular momentum, the magnetic
couplings of different spin states have periodic variations. We argue that
these $L$-dependent spin correlations originate from that the configuration of
localized UHF centers i.e. the classical position of electrons satisfies
$2\pi/N$ rotational symmetry. The gauge invariance of the UHF orbits ensures the phase differences in Eq.(\ref{EQ:Factor}) and results in the
magnetic coupling oscillations. Therefore, we can conclude that these
$L$-dependent spin correlations reflect the nature of spin-dependent RWMs.

\subsection{$S_z$- and $L$-dependent entanglements}
\begin{figure}[ht]
\includegraphics*[width=0.42\textwidth]{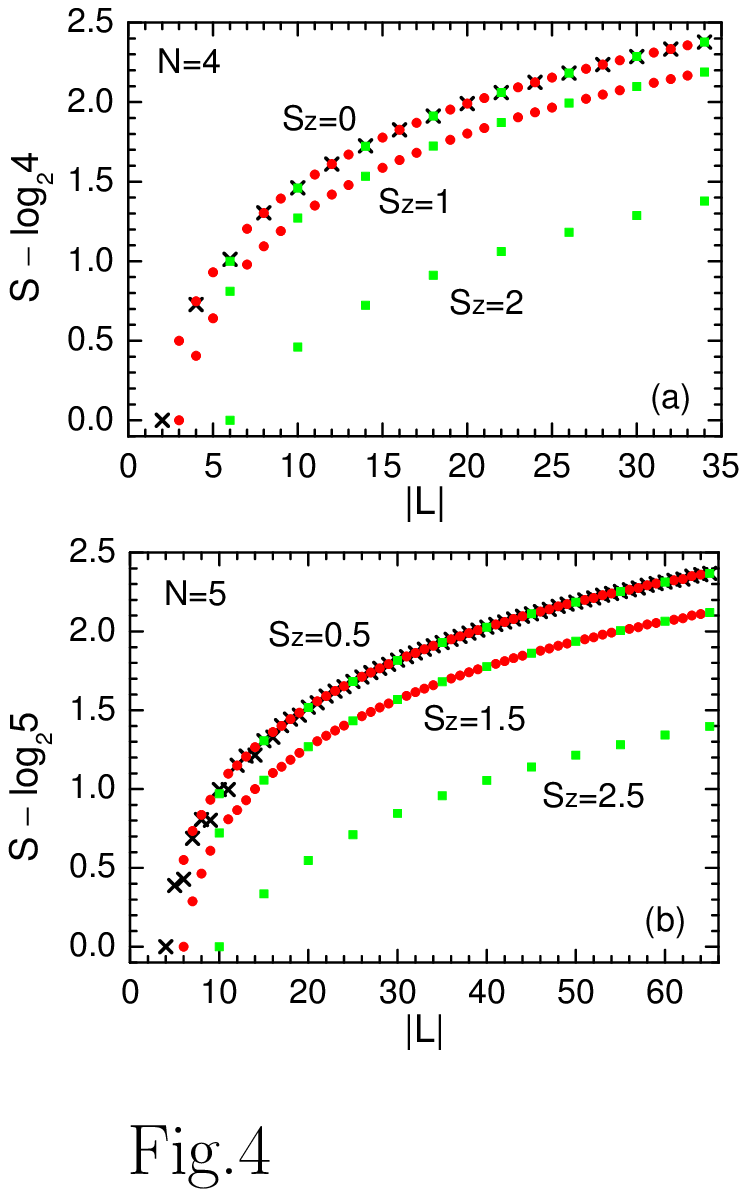}
\caption{\label{FIG:Entropy}(Color online) Entanglement entropies of
the trial wave functions of four (a) and five (b) electrons as
function of angular momentum. $\times$, $\bullet$ and {\Rectsteel}
correspond to the states with non-, partial to full-polarized spins,
respectively. }
\end{figure}
The rotating Wigner molecules are the states driven by strong
particle correlation. When spin degree of freedom is considered, the
correlation is naturally spin-dependent. The magnetic couplings
discussed in the previous subsection are one kind of the reflection
of such spin-dependent correlations.

Now we explore the entanglements of the trial functions which also
reflect the quantum correlations existing in RWMs. The von
Neumann entropy is used for evaluating the 
entanglement between one electron and the other part of the system.
In the case of identical particles, the exchange symmetry of wave
functions contributes a constant $\log_2N$ to the total entropies.
Such parts of the entropies do not reflect the entanglements we 
concern, so we extract them from the total entropies and the
corresponding results are shown in Fig.\ref{FIG:Entropy}, which not
only contain the states with $S_z=0 (0.5)$ but also the ones with
$S_z=1 (1.5)$ and $2 (2.5)$. The entropies given by the trial
functions show that the entanglements of RWMs increase monotonously
with the increase of the angular momentum. For a given $L$, the
states with different $S$ but same $S_z$ have almost equal
entanglement, especially when $|L|$ is large enough. For the states
with same $S$ but different $S_z$, the entropies will have a
constant difference. For the states with largest $S_z$, since all
electrons have same spin, there are no entanglements originating
from spin components and the entropies will differ from that of
the states with smallest $S_z$ by one.

\section{Summary}
In summary, a set of trial wave functions for describing the spin-dependent
rotating Wigner molecular states is constructed from the localized unrestricted
Hartree-Fock orbits with the assumed long-range interaction. The projection
operator technique is employed to restore the rotational symmetry. By examining
the energies of the functions and their overlaps with the exact-diagonalized
ones with the full consideration of the interactions, it is demonstrated that
the trial functions are suitable for the studies of the few-electron states of
quantum dots in strong magnetic fields. The vortex numbers of the functions are
generally $|L|-1$ for the non-full-polarized states with $S_z=0$, where $|L|$
is the angular momentum of the state, except the states with $|L|-1=4n$ whose
vortex numbers are $|L|-2$. They automatically satisfy correct angular momentum
transition rules of different spin states with the increase of the magnetic
fields. These functions explicitly show that there are ferromagnetic and
anti-ferromagnetic couplings between electrons in four-electron rotating Wigner
moleculars, and ferromagnetic and ferrimagnetic couplings in five-electron
case. The different spin states have specific magnetic coupling oscillation
patterns with the change of angular momentum. The allowable angular momenta
and the $L$-dependent magnetic coupling of different spin states are caused by
the $2\pi/N$ rotational symmetry of the classical positions of electrons and
reflect the nature of spin-dependent rotating Wigner moleculars. It is also
shown that the entanglements i.e. the quantum correlations of the states
increase monotonously with the increase of angular momentum. The states
with different $S$ but same $S_z$ have similar entanglement entropies. The
entropy differences of the states with different $S_z$ are approximately
constant in strong magnetic fields. These trial functions will be useful in
future studies of the few-electron states and their quantum behaviors in
quantum dots in strong magnetic fields.

\begin{acknowledgments}
The authors are grateful to Prof. Gun Sang Jeon for the useful information.
Financial supports from NSF China (Grant No. 10574077), the``863" Programme of
China (No. 2006AA03Z0404) and MOST Programme of China (No. 2006CB0L0601) are
gratefully acknowledged.
\end{acknowledgments}

\appendix
\section*{Appendix: Situations for $S_z\neq0(0.5)$}
For four-electron RWMs with $S_z=1$, the many-body bases are
\begin{equation}
\left(
\begin{array}{c}
|1\rangle'\\|2\rangle'\\|3\rangle'\\|4\rangle'
\end{array}
\right)=\left(
\begin{array}{c}
|Z_1^\uparrow,Z_2^\uparrow,Z_3^\uparrow,Z_4^\downarrow\rangle\\
|Z_1^\uparrow,Z_2^\uparrow,Z_4^\uparrow,Z_3^\downarrow\rangle\\
|Z_1^\uparrow,Z_3^\uparrow,Z_4^\uparrow,Z_2^\downarrow\rangle\\
|Z_2^\uparrow,Z_3^\uparrow,Z_4^\uparrow,Z_1^\downarrow\rangle
\end{array}
\right).
\end{equation}
Examining the CPDs or noticing that there is only one spin-down
electron, it can be found that the electrons in these bases have no
specific magnetic coupling.

The corresponding Hamiltonian is
\begin{equation}\label{EQ:Hami43E}
\setlength{\arraycolsep}{-0.5ex}%
H=\left(\hspace{1ex}
    \begin {array}{cccc}
    -2v_1-v_2 &   v_1     &      -v_2 &   v_1    \\
      v_1     & -2v_1-v_2 &   v_1     &      -v_2\\
         -v_2 &   v_1     & -2v_1-v_2 &   v_1    \\
      v_1     &      -v_2 &   v_1     & -2v_1-v_2
    \end {array}\hspace{1ex} \right).
\end{equation}
The eigenvalues and eigenstates of $(H,S^2,S_z)$ are
\begin{equation}\label{EQ:EigenEV43}
\begin{array}{lll}
S=1 & \left\{\begin{array}{l}
E_{1'}=-2v_1\\E_{2'}=-2v_1\\E_{3'}=-2v_2\end{array}\right. &
\begin{array}{l}\psi_{1'}:(-1, 0, 1, 0)\\\psi_{2'}:(0, -1, 0, 1)\\\psi_{3'}:(1,1,1,1)\end{array}\\\\
S=2 & \hspace{2.8ex}E_{4'}=-4v_1-2v_2 &\hspace{0.6ex} \psi_{4'}:(1,
-1, 1, -1).
\end{array}
\end{equation}
It can be found that the states with $S_z=1$ have same energies as
the ones with $S_z=0$ and same $S$.

Again, we expand $|1\rangle'$ to the first Landau level and then the
component with angular momentum $L$ is
\begin{eqnarray}
|1\rangle'_L\!\!&=&\nonumber\\
&&\!\!\sum_{0\leq l_1<l_2<l_3,0\leq
l_4}^{l_1+l_2+l_3+l_4=|L|}{\!\!\left(\prod_{j=1}^4{l_j!}\right)^{-1/2}
\!\!\!\!\!\!\!\!\left(\mathrm{i}^{l_2}-\mathrm{i}^{l_1}\right)
\left(\mathrm{i}^{l_3}-\mathrm{i}^{l_2}\right)}\nonumber\\
&&\!\!\times\left(\mathrm{i}^{l_3}-\mathrm{i}^{l_1}\right)(-\mathrm{i})^{l_4}|l_1^\uparrow,l_2^\uparrow,l_3^\uparrow,l_4^\downarrow\rangle,
\end{eqnarray}
and there are relations as follows
\begin{equation}\label{EQ:Factor43}
\begin{array}{ccc}
|2\rangle'_L&=&\exp(\mathrm{i}6\pi|L|/4)|1\rangle'_L\\
|3\rangle'_L&=&\exp(\mathrm{i}4\pi|L|/4)|1\rangle'_L\\
|4\rangle'_L&=&\exp(\mathrm{i}2\pi|L|/4)|1\rangle'_L
\end{array}.
\end{equation}

Using Eq.(\ref{EQ:Factor43}), four eigenstates in
Eq.(\ref{EQ:EigenEV43}) with restored rotational symmetry can be
obtained. For the states with $S=1$, only those with $|L|=4k\pm1$ or
$|L|=4k$ are allowable. For the states with $S=2$, only those with
$|L|=4k+2$ are nonvanishing. These angular momentum rules are same
as the ones for the states with $S_z=0$. Of course these states do
not have any specific magnetic coupling.

Similarly, it can be obtained that the five-electron RWMs with
$S_z=1.5$ have same energies and angular momentum rules as the ones
with $S_z=0.5$. They can be represented by the many-body basis (up
to a phase factor)
\begin{eqnarray}
&&|Z_1^\uparrow,Z_2^\uparrow,Z_3^\uparrow,Z_4^\uparrow,Z_5^\downarrow\rangle_L\!\!=\nonumber\\
&&\!\!\sum_{0\leq l_1<l_2<l_3<l_4,0\leq
l_5}^{l_1+l_2+l_3+l_4+l_5=|L|}
{\!\!\left(\prod_{j=1}^5{l_j!}\right)^{-1/2}}{\!\!\!\!\!\!\!\!\prod_{i<j}{\left(\delta^{l_j}-\delta^{l_i}\right)}}
\nonumber\\
&&\!\!\times\delta^{4l_5}|l_1^\uparrow,l_2^\uparrow,l_3^\uparrow,l_4^\uparrow,l_5^\downarrow\rangle.
\end{eqnarray}
And they also have no specific magnetic coupling.

Finally, the wave functions for full-polarized RWMs with largest
$S_z$ can be found in other references\cite{Yannouleas2004} and are
no longer discussed here.

\end{document}